# Improved selector behavior in ultrathin chromium-doped $V_2O_3$ films

*Johannes Mohr,[*] Tyler Hennen, Yudi Wang, Xiaoyu Xu, Loc Vinh, Dirk J. Wouters, Rainer Waser, Joyeeta Nag, Daniel Bedau[*]*

J. Mohr, T. Hennen, D. J. Wouters, R. Waser
Institute of Materials in Electrical Engineering and Information Technology II, RWTH Aachen University, 52074 Aachen, Germany
E-mail: mohr@iwe.rwth-aachen.de

Y. Wang, X. Xu, L. Vinh, J. Nag, D. Bedau
Western Digital San Jose Research Center, 5601 Great Oaks Parkway, San Jose, CA 95119, USA
E-mail: daniel.bedau@wdc.com

R. Waser
Peter Grünberg Institute PGI 7, Research Center Jülich, 52428 Jülich, Germany

Funding: Funded in part by the DFG (German Science Foundation) within the collaborative research center SFB 917

**Abstract**
Devices based on the negative differential resistance effect in chromium doped $V_2O_3$ are considered to be promising as selector elements for use in emerging memory technologies, as well as for neuromorphic applications. It is shown by electrical measurements, that the switching effect is maintained for very thin films down to 5 nm, and even improved properties such as a low leakage current and an abrupt transition are observed. For these thicknesses, the behavior of crystalline and amorphous films becomes very similar; most strikingly, a forming step is required in both. Transmission electron microscopy reveals this to be likely due to a thin amorphous layer that forms at the interface to the TiN electrode. Elemental mapping further shows a complex distribution of the chromium dopants, as well as a diffusion of Ti into the layer from the electrode, which might be responsible for the improved properties.

## 1. Introduction

A variety of emerging memory devices, such as those based on ferroelectric (FRAM),[1] magnetic (MRAM)[2] and resistive switching materials (RRAM)[3–5] are beginning to find more widespread use for microelectronic applications. So far, they are primarily employed in niches where their particular characteristics, such as energy efficiency, high density, and relatively easy back end of line (BEOL) integration[6] prove advantageous, as well as in challenging applications such as automotive and space, where temperature stability or radiation hardness are required.[2] In addition, they are under investigation for future computing architectures that attempt to solve the issues posed by the von-Neumann bottleneck. These employ techniques such as computation-in memory[7] or neuromorphic computing.[8] Key to the high-density and especially 3D integration of such emerging memory cells is the availability of a selector device that can be integrated together with the memory element and that does not rely on transistors fabricated during the front end of line process, enabling circuit topologies like 1S1R.[9,10] An ideal selector is completely insulating below its threshold voltage, above which it turns into a low resistive conductor. They need to have a high selectivity, i.e. a high ratio between off- and on-state resistances, a low leakage current, narrow distributions and process compatibility with the memory element.[11,12]

One of the critical requirements for a selector material is the ability to function effectively even at very small thicknesses. Especially for the case of 3D integration, where selector planes will be perpendicular to the wafer surface, the selector thickness directly affects the achievable density in X and Y direction.

One material that has shown promise for integrated selector devices is chromium doped $V_2O_3$ ($Cr:V_2O_3$). It is thought to be a prototypical Mott-insulator[13], and it has been proposed that an electronically triggered metal-insulator transition could be the origin of the switching phenomenon,[14] from which devices showing a negative differential resistance (NDR) behavior can be fabricated with very high endurance and fast switching.[15,16] While it has been shown that the transition can be observed even in thin-films with significant imperfections,[17] the competing explanation of a thermal runaway effect appears more plausible due to the excellent agreement with the measurement data.[15,16,18] A key limitation so far have been the relatively thick films used for most studies, commonly on the order of 30 nm. This is an issue for the integration into high density memories, which typically feature films thinner than 10 nm due to the requirements of lithography and film deposition. Therefore, it is necessary to investigate whether the Cr:$V_2O_3$ thickness can be reduced while maintaining satisfactory switching performance, avoiding for example increased leakage currents. Here, devices with 5 nm film thickness were fabricated and interestingly show much better behavior than anticipated. To elucidate the origin of this, a comprehensive investigation by high-resolution transmission electron microscopy (TEM) and nanoscale elemental mapping was conducted.

## 2. Experimental

The devices used in this work were based on TiN bottom point contact nano-via structures as shown in **Figure 1**. The vias are embedded into an insulating layer, with a negligible step height. The shape of the via as seen from the top is a rounded square; different widths, and correspondingly device sizes, between 120 nm and 500 nm are available. From below, the vias are contacted by a metal line, that connects to a pad. Devices with a direct connection are available, as well ones with an integrated resistor. On top of the via is the active Cr:$V_2O_3$ layer, which is capped by a platinum top electrode, that serves both to connect another pad to the device and to protect it from ambient influences such as moisture and oxidation. Top and bottom electrodes are isolated by an undercut mesa structure. Both Cr:$V_2O_3$ and the top electrode extend laterally beyond the via; the active device volume is defined by the area of the via. These structures have been used previously to investigate the NDR in thicker Cr:$V_2O_3$ films,[15] a TEM cross-section is shown in the reference.

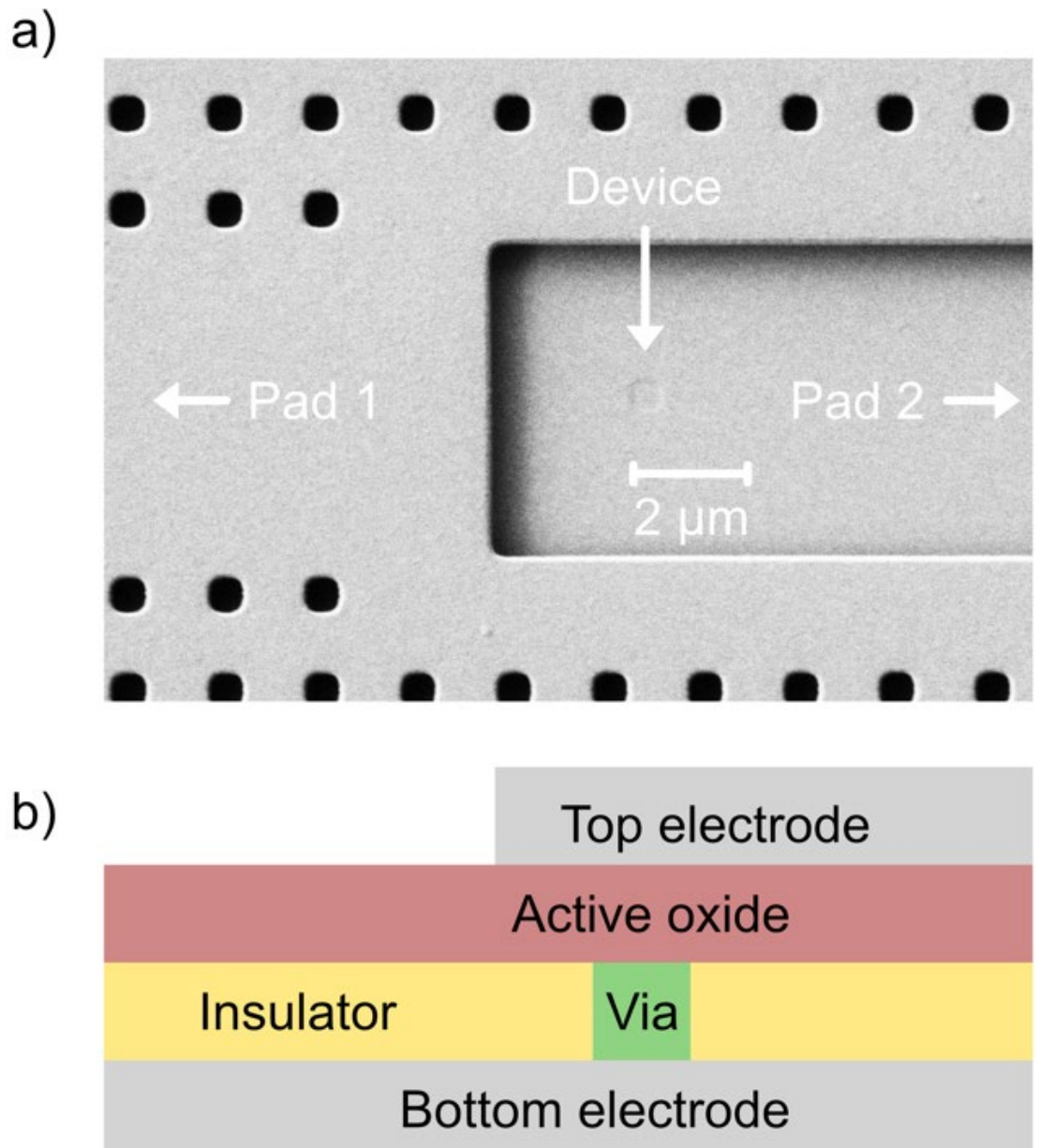

**Figure 1:** a) SEM image of the finished structure. The device is visible in the center as a small depression in the top platinum electrode. b) Sketch of the cross-section of the stack along a horizontal line in the center of a).

The devices were created by blanket deposition of material on top of coupons diced from a nano-via vehicle wafer. The coupons were cleaned in an ultrasonic bath for 10 min in acetone and isopropanol, followed by a rinse in deionized water, drying with nitrogen gas and a sputter etching step, to remove the oxidized top surface of the via. An acceleration voltage of 1000 V and a duration of 17 s were used. To prevent any formation of new oxide the samples were then immediately transferred into the vacuum of the deposition system. Next, the $Cr:V_2O_3$ films were grown by reactive radio-frequency (RF) magnetron sputtering from a 1” target consisting of an alloy of 15% Cr and 85% V. A gas mixture consisting of argon with 600 ppm oxygen mixed in was introduced into the chamber to provide an oxidizing atmosphere. The low oxygen content is necessary to prevent the formation of higher vanadium oxides.[19] The pressure in the chamber was automatically regulated to a value of 0.01 mbar. Heating of the substrates to a deposition temperature of 600°C led to the growth of the desired crystalline corundum type phase. This has been previously verified by blanket depositions and grazing incidence Xray diffraction (GIXRD) measurements.[19] The film thickness was controlled by the deposition time, which was calibrated by X-Ray reflectivity (XRR) measurements to yield 5 nm thick layers. After the deposition, the samples cooled inside the chamber to a temperature of about 100°C, because it was found that for undoped $V_2O_3$, the properties depend slightly on the temperature at which the samples are removed.[20] In addition, X-Ray amorphous samples were prepared by depositions at room temperature. Then, without breaking the vacuum, a 30 nm layer of platinum was deposited by RF sputtering in an inert atmosphere to form the top electrode.

In addition to the devices, another set of samples was prepared for a TEM investigation. They consisted of blanket films grown on silicon wafers covered with a 5 nm TiN bottom electrode. The cleaning and film deposition were done identically to the devices. However, due to the very thin bottom electrode, no ion-cleaning step was performed here. As the top electrode only serves to prevent oxidation in this case, it was chosen to be 10 nm thick. Due to the experimental effort and relevance for typical devices, only crystalline films were investigated.

For the electrical characterization, the devices were contacted using a custom-built probing setup. The pristine resistance was measured using a Keithley 2636B source-meter unit (SMU) directly on a device. The NDR on the other hand was characterized using a combination of a waveform generator and an oscilloscope on devices with an integrated series resistor of approximately 8.2 kΩ. The measurements were done using voltage sweeps of different amplitudes while recording the current utilizing a python package specifically developed for the measurement and analysis of NDR devices.[21] The duration of the sweeps was 1 ms or less for the NDR measurements and significantly longer for the resistivity measurements. To derive the actual device behavior, the voltage drop over the series resistor can be calculated and subtracted from the applied voltage $V_{\mathrm{app}}$ to yield the device voltage $V_{\mathrm{D}}$:

$$V_{\mathrm{D}} = V_{\mathrm{app}} - R_{\mathrm{series}}\, I$$

The resistivity of the material was characterized by measuring a number of devices using voltage sweeps with an amplitude of 100 mV to avoid any switching effects. A line was fit to the I-V data to extract the resistance.

For the TEM measurements, lamellas were prepared from the films by focused ion beam (FIB) using a Thermo Fisher Scientific Helios 600 FIB, so that cross-sectional images could be taken. For this, they were first covered with a protective $SiO_2$ layer in the FIB on top of the platinum. Imaging was then done using a JEOL ARM-200CF electron microscope, in addition, elemental maps were recorded using a combination of Electron Energy Loss Spectroscopy (EELS) and Energy-dispersive X-ray spectroscopy (EDX).

## 3. Results and Discussion

Examples of the switching behavior for both crystalline and amorphous devices are shown in **Figure 2**. In either case, initially there is a forming event, which leads to an abrupt increase in the current when a certain threshold voltage is exceeded. This forming step permanently alters the film's behavior. In successive loops, a strongly non-linear behavior is observed. Initially, below 1.0 V to 1.5 V, the leakage current is very small. Even at 1 V, it is only 1.4 μA for the crystalline device and 7.5 μA for the amorphous one, which appears very promising for applications: In a large array, the leakage currents through many not selected cells add up, and it is therefore necessary for each individual one to leak as little as possible, both for reliable operation and to reduce energy consumption. If the threshold voltage is exceeded, the current increases rapidly over a very narrow range on the order of 10 mV. This leads to the appearance of a nearly vertical characteristic. Upon closer inspection, especially for the amorphous device, the I-V data show the typical NDR behavior that has been previously observed in both crystalline[15,16,18] and amorphous devices.[22] Thus, the devices are a very good approximation to an ideal threshold switch, that exhibits either infinite or zero resistance, depending on whether the applied voltage is below or above the threshold.

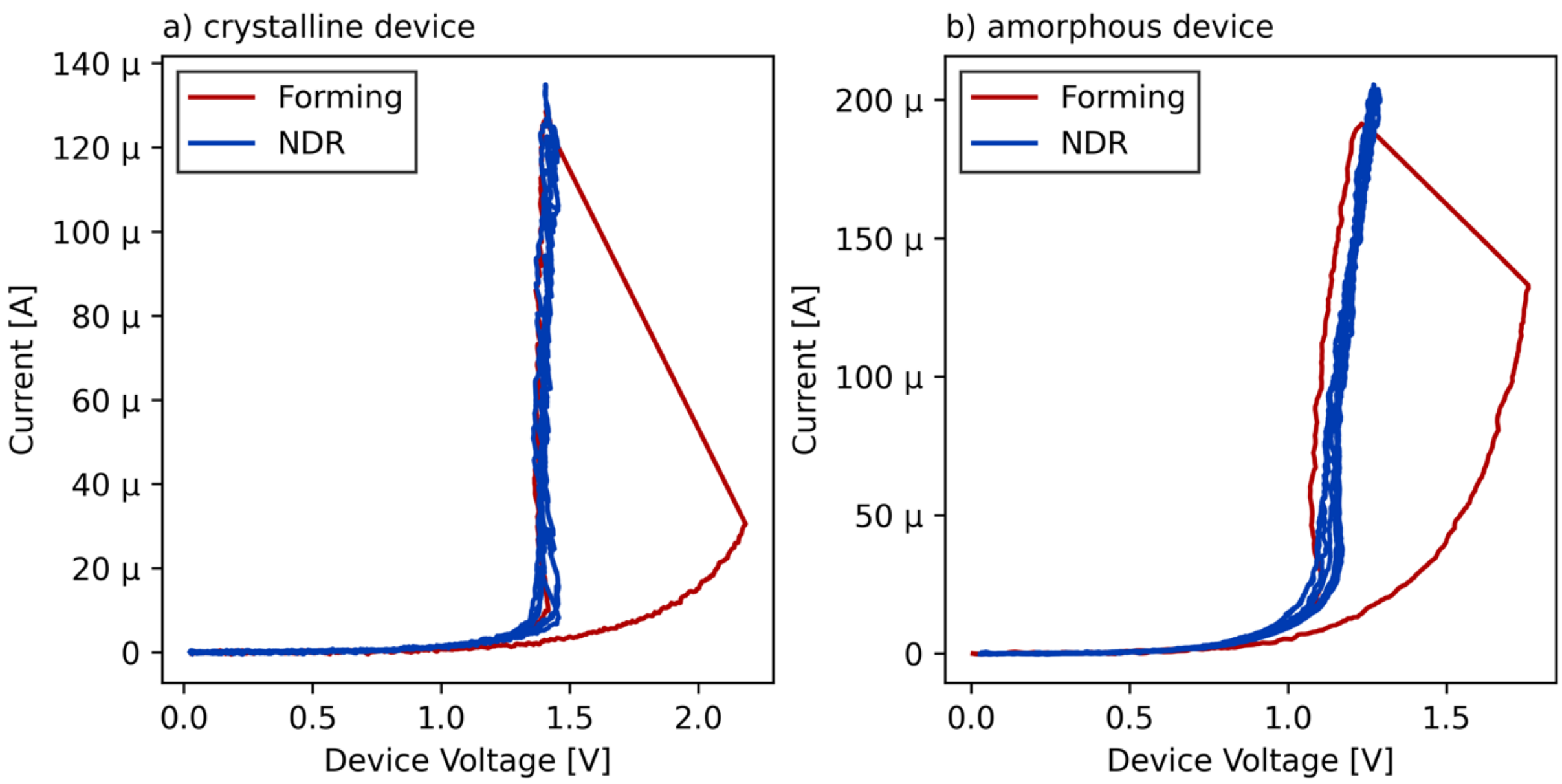

**Figure 2**: Negative differential resistance behavior (blue) is observed after an initial forming loop (red) for both the crystalline and amorphous device. The shown cells were 150 nm wide for a) and 250 nm for b).

Interestingly, in both cases a forming step is necessary. For amorphous films this is expected, because it has been previously observed for thicker films.[22,23] On the other hand, crystalline films do not usually require forming.[15] The need for a forming step is only slightly inconvenient for a practical fabrication process, because the devices need to be contacted for an electrical test anyways; but a forming step might prove difficult to control well enough for predictable device properties. The very similar electrical behavior of course suggests that the underlying materials are also comparable, for example, the crystallinity of the films might be reduced for very thin layers. This is further confirmed by the pristine resistivities of the cells shown in **Figure 3**.

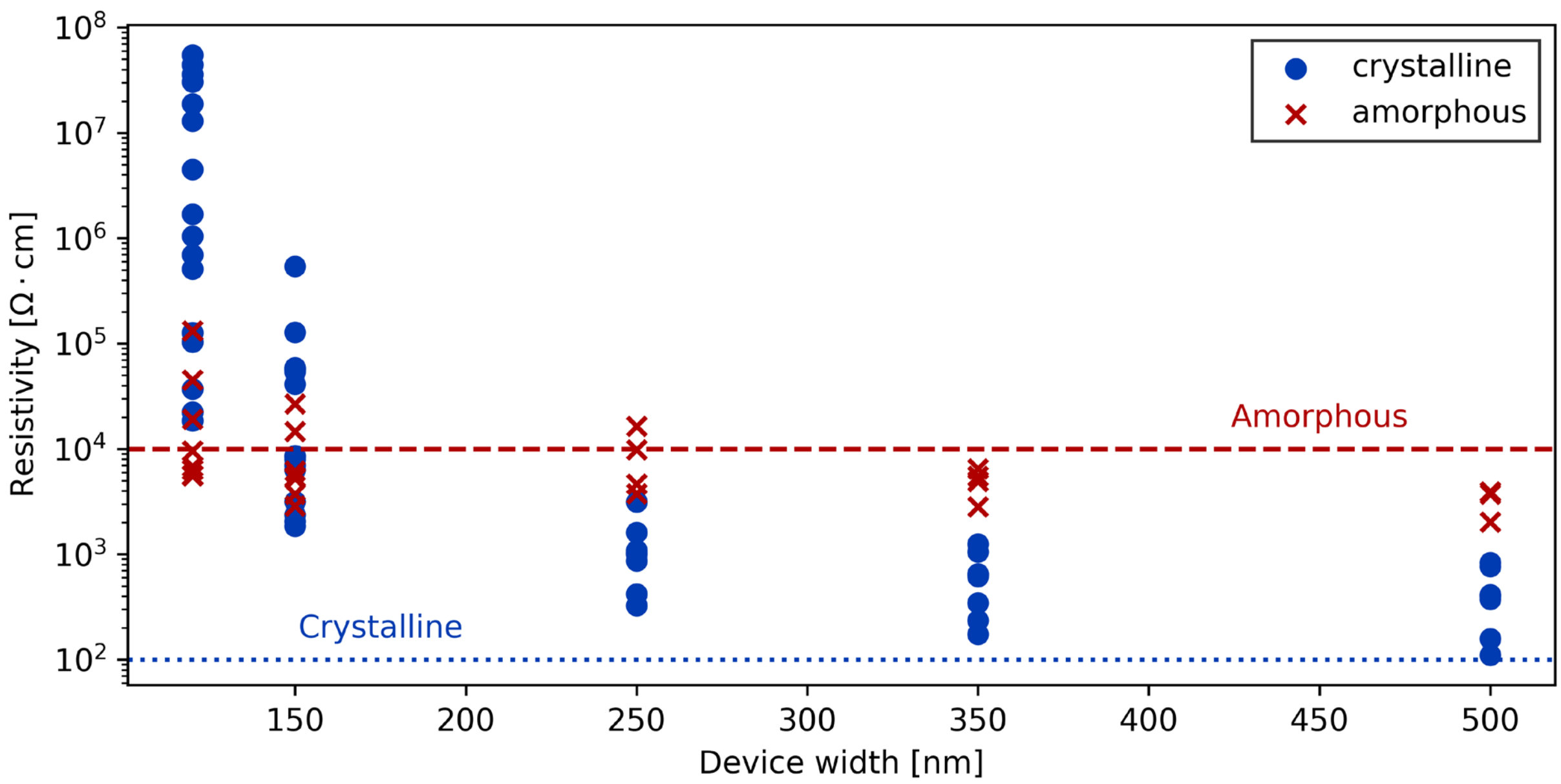

**Figure 3**: Resistivities calculated for the different cell sizes assuming a homogeneous film over the device width and film thickness. Horizontal lines indicate approximate values for the expected bulk resistivity for single crystals[13] and thick amorphous[18] films.

For any device width, the resistivities differ much less between crystalline and amorphous than the expected two orders of magnitude. For large cells, crystalline devices show resistivities close to, but slightly higher than the literature values. As device width decreases, the resistivity increases, but only slightly for the amorphous material, whereas it is increased by more than a factor of 1000 for the crystalline case. Because of this, for very small cells, the crystalline devices are actually more resistive, which appears to be consistent with the lower leakage current observed in the switching measurements.

In the amorphous case, the behavior appears plausible. For small cells, the resistivity is quite close to thick films from the literature, indicating that they are probably not fundamentally different. For larger ones, defects in the films that are expected to be more significant in these very thin layers become more likely to affect a device, resulting in a lower effective resistivity. On the other hand, for the crystalline material, the behavior is quite difficult to understand. Even if thin layers had a reduced crystallinity, it is not clear how the resistance can be higher than for fully amorphous ones, and furthermore why this would depend on the device size. To clarify this, we carried out high-resolution TEM imaging shown in Figure 4.

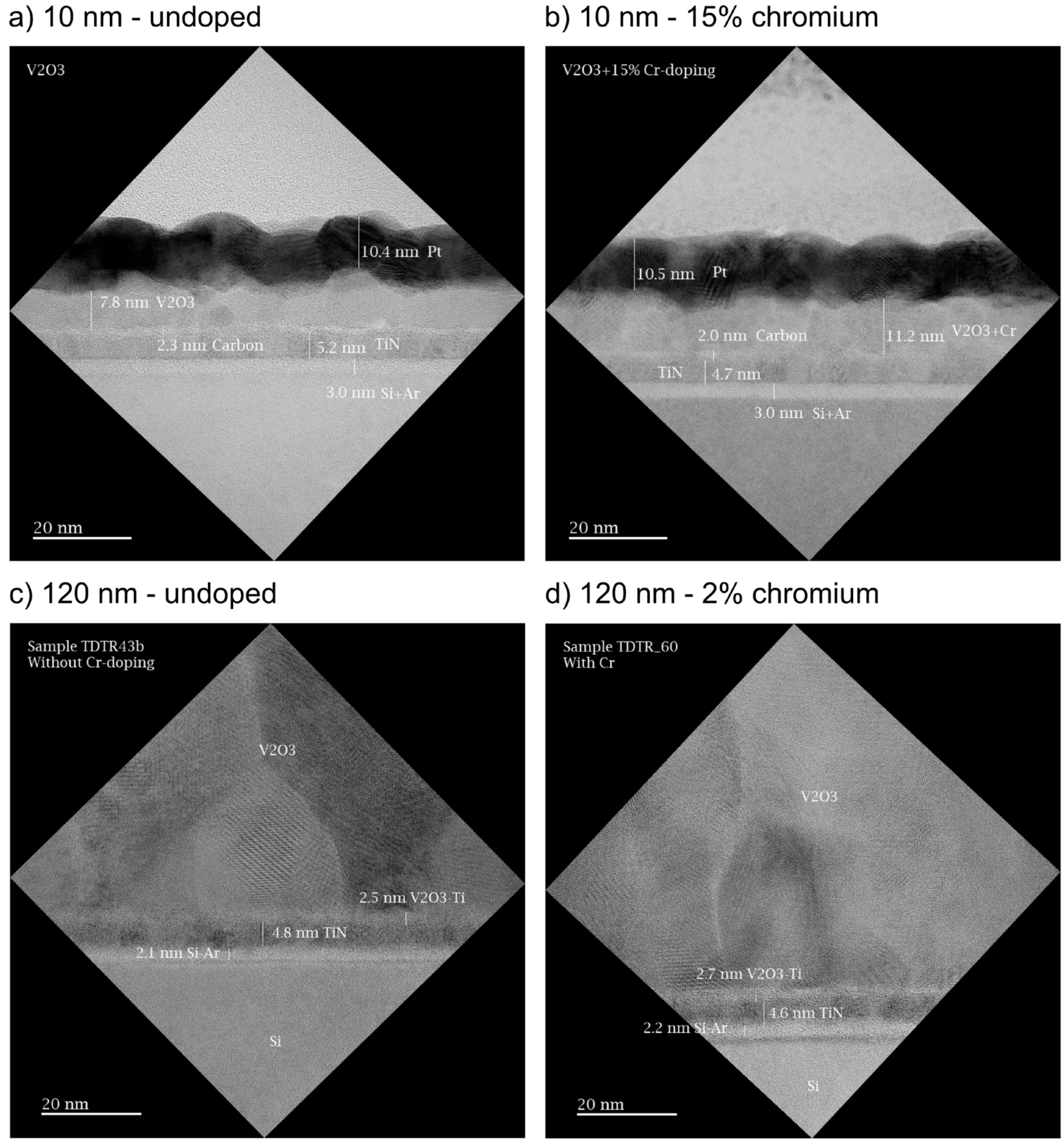

**Figure 4**: Bright field TEM images of four types of $V_2O_3$ films deposited on TiN bottom electrodes capped with Pt: The top row shows 10 nm thick layers, the bottom one 120 nm. The left images represent undoped $V_2O_3$, whereas on the right they are chromium doped. Note that the doping concentration is different, this was done to investigate whether it has a significant influence on the structure.

**Figure 4** demonstrates that the films show an approximately columnar growth mode, with grains that are about 5 nm to 10 nm in size for the 10 nm films, while for 120 nm they are significantly bigger. As these images were taken at the same magnification for comparability, the thick films are not visible completely. They are shown at a reduced magnification in **Figure S1**, which demonstrates that the grains are on the order of 20 nm for the 2% doped film, and approximately 50 nm for the undoped one.

The latter also has a more disordered grain structure. This indicates that for the 5 nm layers used in the devices, even those with the smallest width of 120 nm will consist of many grains, and therefore differences in the precise grain structure should not lead to a large variability between devices in even further scaled cells. At this thickness, the films are also quite smooth, and thickness fluctuations should contribute little variability. On the other hand, for thick films, and especially the undoped ones, a significant surface roughness is observed (**Figure S1**). This indicates that as expected, thin layers are preferable for applications.

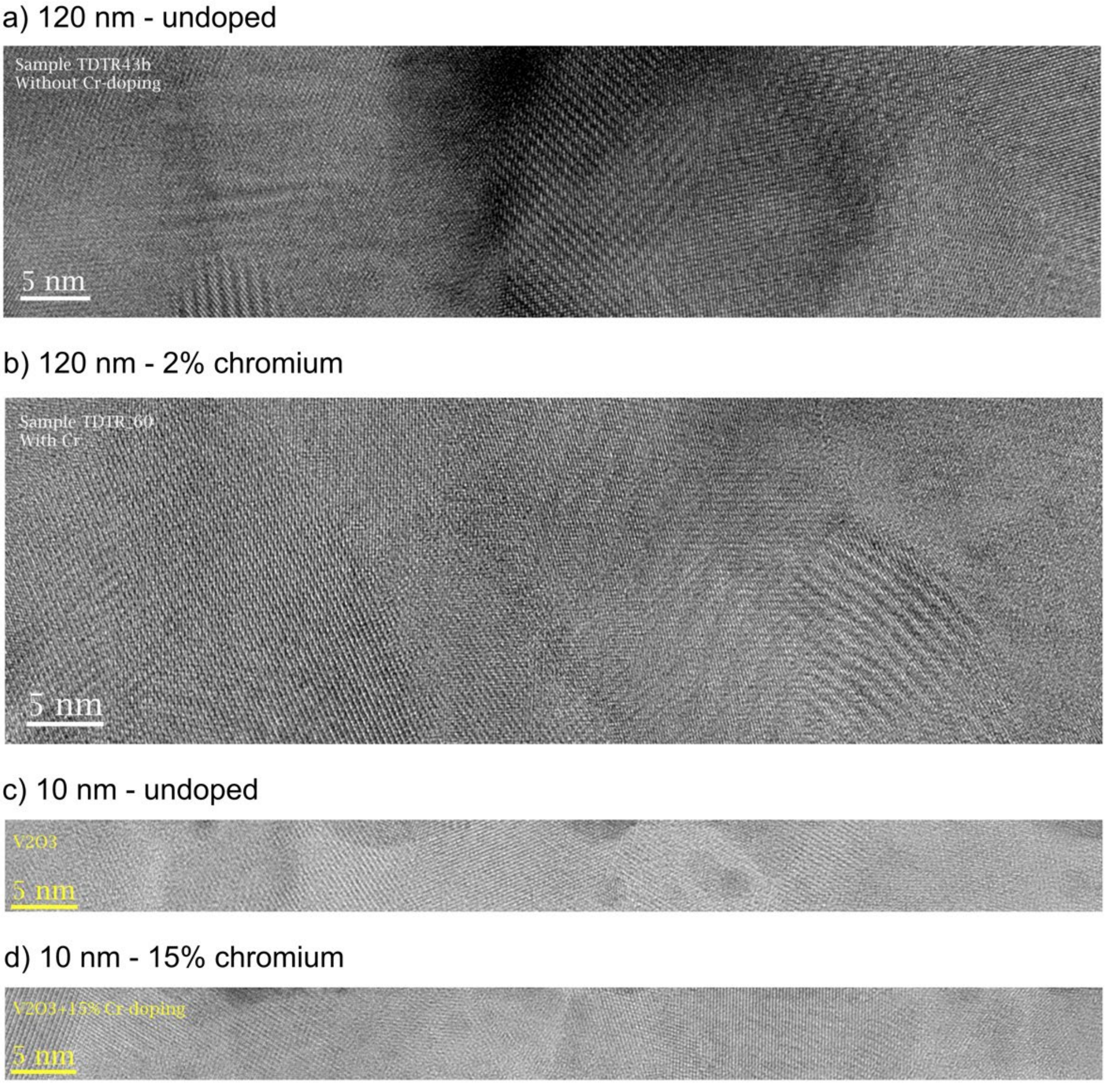

**Figure 5**: Higher magnification images of the films shown in Figure 4.

To assess the crystallinity of the films, higher magnification images were acquired, they are shown in **Figure 5**. Clearly, even at 10 nm thickness, the bulk remains well crystallized, indicating that this not the origin of behavior similar to amorphous films at reduced thicknesses. However, Figure 4 shows an interesting feature at the bottom interface of the films. Between the TiN electrode and the $V_2O_3$ film, an intermediate interfacial layer forms, which appears amorphous. It is approximately 2-3 nm thick, regardless of the film thickness and doping concentration. Of course, such a layer could lead to an amorphous-like switching behavior in very thin, nominally crystalline devices, because it makes up a significant portion of the 5 nm film. However, it must also be considered what this layer is composed of. To elucidate this, EELS/EDX maps are shown in **Figure 6** for the 10 nm films; very similar results are found for 120 nm.

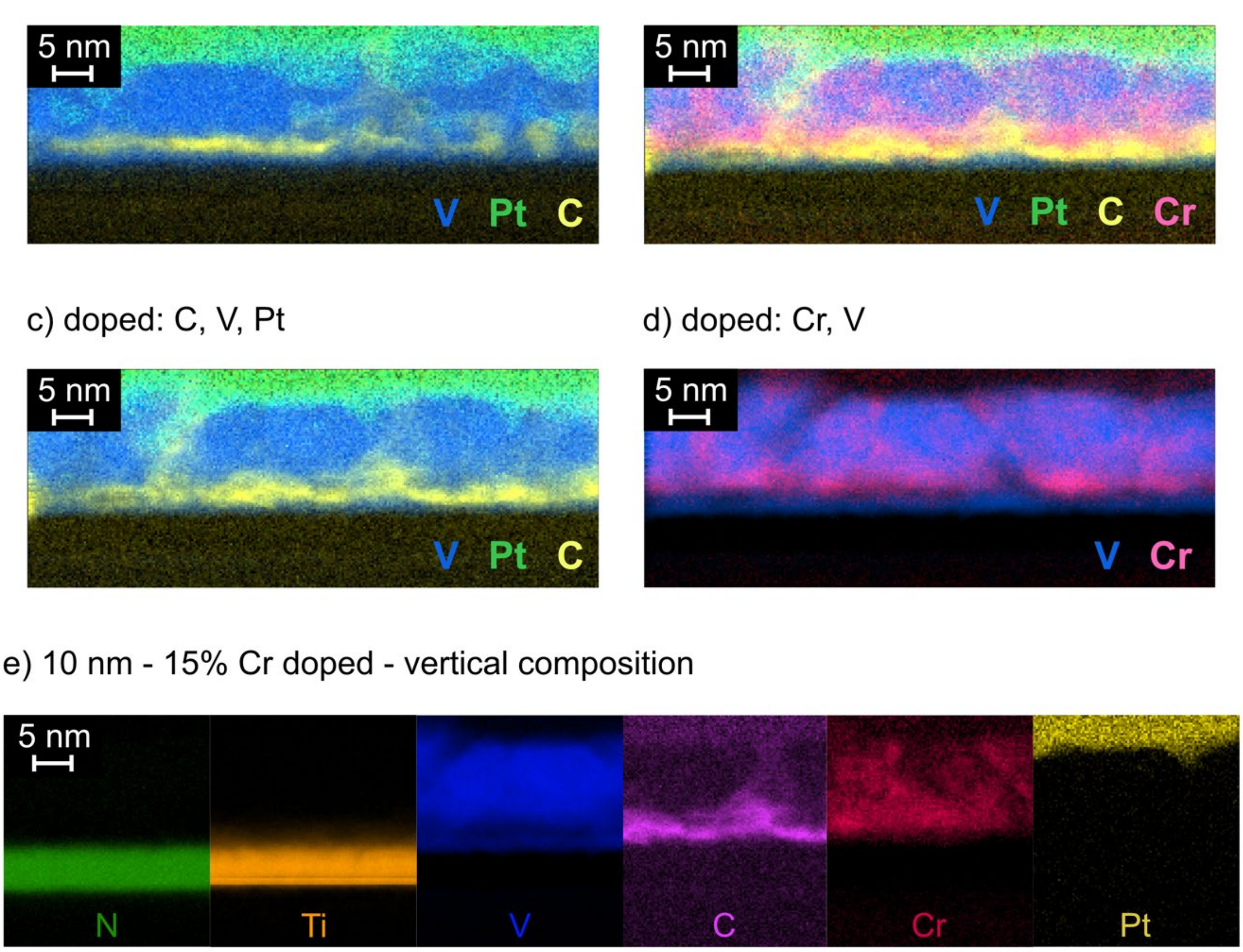

**Figure 6**: a-d) EELS/EDX maps of undoped and 15% Cr doped 10 nm $V_2O_3$ films. For clarity, the elements are overlayed in different combinations for the doped case. To clearly show the structure of the bottom interface, the signals are presented next to each other in e).

Clearly, the composition is inhomogeneous, both in the plane of the films and along the thickness direction, but not all features are likely to have a significant influence on the electrical device behavior. For example, a carbon contamination layer at the bottom interface is observed, that also permeates the grain boundaries to a certain degree (Figure 6 a, c)). This contamination layer is most likely not present in the bottom point contact devices, which received an additional ion etching step before film deposition. Additionally, while it appears like a solid layer due to the scaling of the signal, the actual concentration of carbon is only on the order of 5 at%. Therefore, it is unlikely to have a large influence on the electrical observations. The Pt top electrode also slightly penetrates into the grain boundaries, which might reduce the effective film thickness somewhat. Interestingly, contrary to what has been previously suspected, there is no indication of a segregation of chromium there (Figure 6 d)).

On the other hand, on the bottom interface, a layered structure is found that might explain the device properties (Figure 6 d)): Starting with the TiN electrode, the first layer is Cr poor, followed by a Cr rich one, and finally by an average bulk stoichiometry. Especially curious is the clear separation of the two layers. As distinguishing multiple elements is difficult in the overlaid maps, the detailed stack is shown separately in Figure 6 e). The TiN bottom electrode in principle appears to have a sharp top interface, at which the Cr:$V_2O_3$ layer starts. However, a clear diffusion of Ti into this is also visible, which is not surprising given the high growth temperature of 600°C. The V concentration in this interdiffusion layer appears to be slightly lower than in the rest of the film and it seems to correlate with the Cr poor region. Judging by the thickness, it should furthermore correspond to the amorphous layer seen in Figure 4. This is then followed by the Cr rich layer. The carbon contamination appears to extend largely through the whole bottom structure.

Combining these findings with the electrical data, the NDR behavior in very thin Cr:$V_2O_3$ devices can be understood. Likely it always occurs in an amorphous layer, either a 2-3 nm thick interfacial layer in the nominally crystalline films or the whole 5 nm film in the amorphous ones. Probably, the mechanism is the same as observed in thicker amorphous films, where in an initial forming step a crystallized filament is formed, in which then a thermal runaway can lead to NDR behavior.[22,23] While for the amorphous devices the filament would be homogeneously Cr-doped and through the whole layer, in the crystalline ones, a Ti-doped filament through the interfacial layer might form, followed by the fully crystalline, highly Cr-doped layer. The latter would then resemble a scaled down version of the bottom point contact structure of the devices, especially as the filament should be relatively conductive due to the Ti doping.[24]

This might be an explanation for the surprisingly good switching of the crystalline devices. From the known resistivity of $Cr:V_2O_3$ single crystals,[13] a leakage current orders of magnitude higher than for the amorphous devices would be expected (Figure 2), and threshold switching would be unlikely to occur. In reality, however, the leakage is lower, and the threshold more abrupt. This is probably related to the pristine resistivity of the cells, which is higher than expected for all widths and increases with a decrease in device size (Figure 2). The former is readily explained by the Cr rich layer observed in TEM, because the resistivity of $Cr:V_2O_3$ increases with doping concentration.[13] Thus, a highly resistive layer is formed that likely dominates the total resistance.

The scaling with device size can be understood due to the interfacial amorphous layer. For large devices, there is a high likelihood of a defect breaking this layer, and thus the resistivity of the Cr-rich layer is measured. For small ones however, the interface will dominate the resistivity, which thus approaches that of amorphous devices, and even exceeds it for 120 nm cells (Figure 2). The origin of the latter is not quite clear, but it seems plausible that the Ti doping could affect the resistivity; while this effect is well understood for crystalline materials,[24,25] unfortunately, no literature data on the influence of Ti on amorphous $V_2O_3$ appear available for comparison.

Considering the effective device size, which is likely smaller than the cell and controlled by the filament through the interfacial layer, the observed behavior appears plausible: Extrapolating data from 15% doped crystalline devices, for a 10 nm thick and 40 nm wide cell, a threshold voltage of slightly more than 1.5 V and a leakage current of 1 µA at 1 V were calculated.[15] Considering that a lower layer thickness will reduce the threshold voltage, whereas a smaller width will increase it, the present data seem to be in a good agreement with the expectation for a crystallized filament a few tens of nanometers wide. The very sharp transition might be due to the effectively higher Cr doping concentration.

## 4. Conclusion

The results of this study demonstrate that it is not only possible to reduce the thickness of the active layer in Cr:$V_2O_3$ selector devices down to 5 nm, but that the selector properties are also enhanced, with significantly reduced leakage currents (~100 nA for voltages < 0.5 V) and a steeper transition. However, unlike thicker films, even in the crystalline material, a forming step is required to activate the selector behavior. This behavior, resembling that of amorphous films, is believed to result from switching occurring in a thin amorphous interfacial layer. This suggests that the layer thickness could potentially be further reduced to 2-3 nm. These findings show promising potential for applications, as an amorphous film of the required thickness could be directly deposited, avoiding the high temperatures necessary for crystallization, which hinder the integration with other CMOS electronics. Because the forming voltages are on the order of 2 V, and less than 1 V higher than the threshold voltages, they do not impose significant limitations on the surrounding circuits, and the forming step could be performed during the electrical testing of the devices. The origin of the improved selector characteristics is not quite clear however, they are likely due to a specific doping profile in the films, potentially involving both chromium and titanium doping. To elucidate this, further experiments with intentionally varied profiles are necessary, which should be done with amorphous films to avoid interdiffusion during the deposition.

## Acknowledgements

JM, TH, DJW and RW gratefully acknowledge funding by the DFG (German Science Foundation) within the collaborative research center SFB 917.

## Data Availability Statement

The data that support the findings of this study are available from the corresponding author upon reasonable request.

## Conflict of Interest Statement

The authors have no conflicts to disclose.

## References

[1] Volker Rzehak, *Low-Power FRAM Microcontrollers and Their Applications*, Texas Instruments.

[2] J. Heidecker, *MRAM Technology Status*, Pasadena, California, 2013.

[3] Efficient and high-performing vehicle architecture: Cooperation of Continental and Infineon, 2023.

[4] Qualcomm QCC730 Dual band micro-power Wi-Fi Product Brief, .

[5] Qualcomm QCC711 Tri-core Ultra-Low Power Bluetooth Low Energy SoC Product Brief, .

[6] H. Lv, X. Xu, P. Yuan, D. Dong, T. Gong, J. Liu, Z. Yu, P. Huang, K. Zhang, C. Huo, C. Chen, Y. Xie, Q. Luo, S. Long, Q. Liu, J. Kang, D. Yang, S. Yin, S. Chiu, M. Liu, In *2017 IEEE International Electron Devices Meeting (IEDM)*, IEEE, 2017, pp. 2.4.1-2.4.4.

[7] A. Sebastian, M. Le Gallo, R. Khaddam-Aljameh, E. Eleftheriou, *Nat Nanotechnol* 2020, *15*, 529.

[8] Y. Li, Z. Wang, R. Midya, Q. Xia, J. J. Yang, *J Phys D Appl Phys* 2018, *51*, 503002.

[9] Y. Cho, J. Heo, S. Kim, S. Kim, *Surfaces and Interfaces* 2023, *41*, 103273.

[10] E. Cha, J. Woo, D. Lee, S. Lee, J. Song, Y. Koo, J. Lee, C. G. Park, M. Y. Yang, K. Kamiya, K. Shiraishi, B. Magyari-Kope, Y. Nishi, H. Hwang, In *2013 IEEE International Electron Devices Meeting*, IEEE, 2013, pp. 10.5.1-10.5.4.

[11] S. A. Chekol, J. Song, J. Park, J. Yoo, S. Lim, H. Hwang, In *Memristive Devices for Brain-Inspired Computing*, Elsevier, 2020, pp. 135–164.

[12] X. Peng, R. Madler, P.-Y. Chen, S. Yu, *J Comput Electron* 2017, *16*, 1167.

[13] D. B. McWhan, J. P. Remeika, *Phys Rev B* 1970, *2*, 3734.

[14] P. Stoliar, L. Cario, E. Janod, B. Corraze, C. Guillot-Deudon, S. Salmon-Bourmand, V. Guiot, J. Tranchant, M. Rozenberg, *Advanced Materials* 2013, *25*, 3222.

[15] T. Hennen, D. Bedau, J. A. J. Rupp, C. Funck, S. Menzel, M. Grobis, R. Waser, D. J. Wouters, In *2018 IEEE International Electron Devices Meeting (IEDM), 1-5 December 2018*, San Francisco, CA, USA, 2018, pp. 37.5.1-37.5.4.

[16] T. Hennen, D. Bedau, J. A. J. Rupp, C. Funck, S. Menzel, M. Grobis, R. Waser, D. J. Wouters, In *2019 IEEE 11th International Memory Workshop (IMW)*, Monterey, CA, USA, 2019, pp. 1–4.

[17] J. Mohr, T. Hennen, D. Bedau, R. Waser, D. J. Wouters, *Advanced Physics Research* 2024, *3*, 2400040.

[18] T. Hennen, Harnessing stochasticity and negative differential resistance for unconventional computation. Dissertation, RWTH Aachen University, 2023.

[19] J. A. J. Rupp, Synthesis and Resistive Switching Mechanisms of Mott Insulators based on Undoped and Cr-doped Vanadium Oxide Thin Films. Dissertation, RWTH Aachen University, 2020.

[20] J. Trastoy, Y. Kalcheim, J. del Valle, I. Valmianski, I. K. Schuller, *J Mater Sci* 2018, *53*, 9131.

[21] Tyler Hennen, *thennen/py-ivtools*, Zenodo, 2024.

[22] J. Mohr, C. Bengel, T. Hennen, D. Bedau, S. Menzel, R. Waser, D. J. Wouters, *physica status solidi (a)* 2024, *221*, 2300405.

[23] J. A. J. Rupp, M. Querré, A. Kindsmüller, M.-P. Besland, E. Janod, R. Dittmann, R. Waser, D. J. Wouters, *J Appl Phys* 2018, *123*, 44502.

[24] D. B. McWhan, A. Menth, J. P. Remeika, W. F. Brinkman, T. M. Rice, *Phys Rev B* 1973, *7*, 1920.

[25] H. KUWAMOTO, H. V. KEER, J. E. KEEM, S. A. SHIVASHANKAR, L. L. VAN ZANDT, J. M. HONIG, *Le Journal de Physique Colloques* 1976, *37*, C4.

# Supporting Information

a) 120 nm - undoped b) 120 nm - 2% chromium

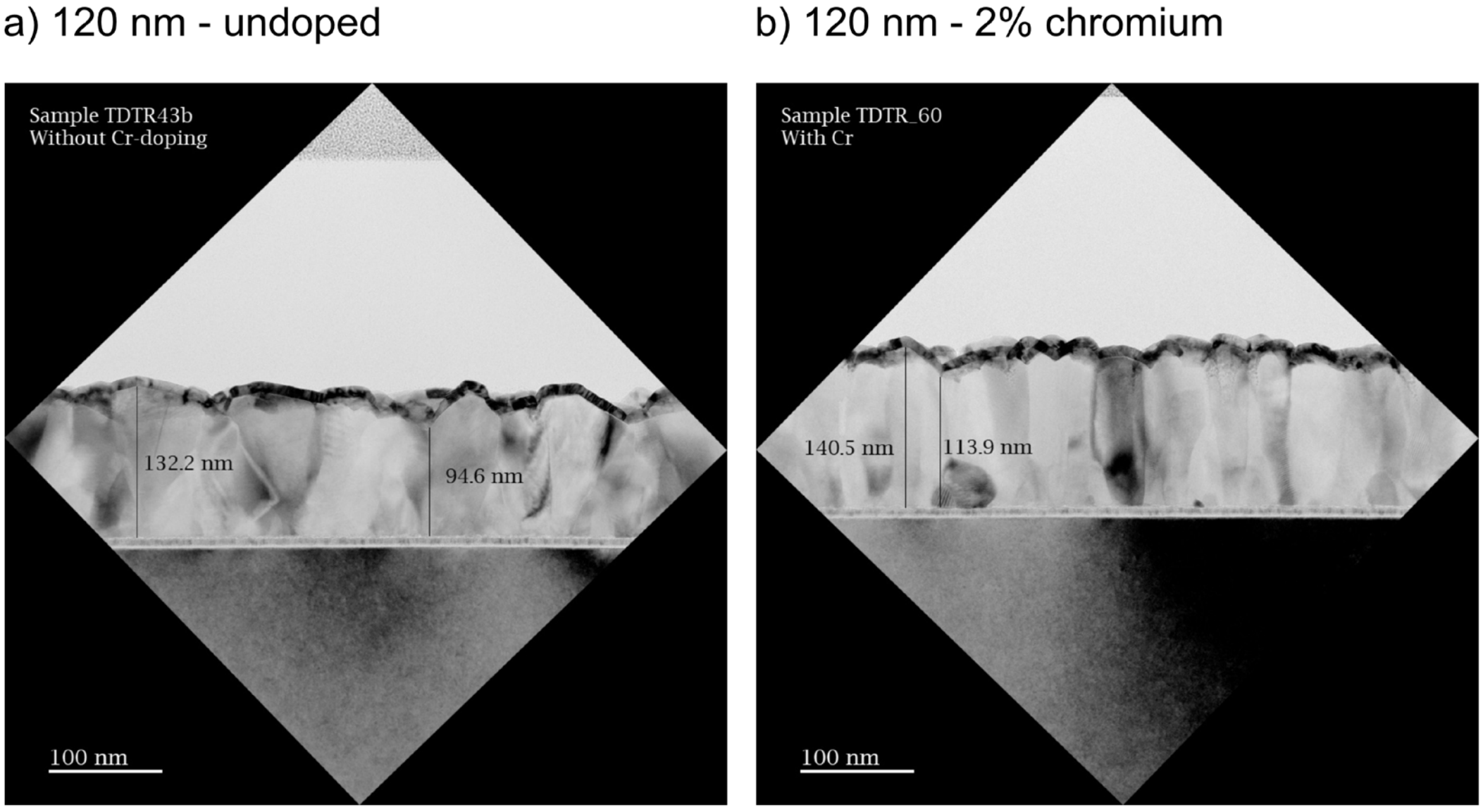

**Figure S1:** Lower magnification view of the films shown in Figure 4.